\documentclass{article}
\usepackage{iclr2027_conference,times}
\usepackage{amsmath,amsfonts,bm}

\def\eqref#1{equation~\ref{#1}}
\def\1{\bm{1}}

\DeclareMathAlphabet{\mathsfit}{\encodingdefault}{\sfdefault}{m}{sl}
\SetMathAlphabet{\mathsfit}{bold}{\encodingdefault}{\sfdefault}{bx}{n}

\usepackage{hyperref}
\usepackage{url}
\usepackage{booktabs}
\usepackage{graphicx}
\usepackage{float}
\title{MotionPlan: Training-Free LLM-Guided Residue Motion Planning for Protein Dynamics Generation}
\iclrfinalcopy
\author{\begin{tabular}{c}
Chenyu Liu\textsuperscript{1,2,4},
Bowen Gao\textsuperscript{1,2,4},
Haichuan Tan\textsuperscript{1,2,4},
Jianhui Wang\textsuperscript{1,3},
Yanyan Lan\textsuperscript{1,2,\(\dagger\)}\\[6pt]
\small \textsuperscript{1}Institute for AI Industry Research (AIR), Tsinghua University\\
\small \textsuperscript{2}AIPher Technology\\
\small \textsuperscript{3}University of Electronic Science and Technology of China\\
\small \textsuperscript{4}Department of Computer Science and Technology, Tsinghua University\\[4pt]
\small \textsuperscript{\(\dagger\)}Corresponding author: \texttt{lanyanyan@air.tsinghua.edu.cn}
\end{tabular}}
\begin{document}

\maketitle
\begin{abstract}
Protein dynamics generators can produce plausible states and temporally ordered rollouts, yet their objectives largely fit marginal distributions rather than the inter-residue coordination organizing transitions; coordination errors therefore accumulate over long horizons. Retraining for explicit residue-level control is costly and requires paired motion supervision rarely available in MD data. We introduce \emph{MotionPlan}, a training-free framework in which a large language model (LLM) infers per-residue coordination roles and confidence from sequence and structural context. The plan becomes a localized, confidence-weighted intervention on a frozen generator's pair representations throughout reverse-SDE sampling, without parameter updates or coordinate editing. Because Pairformer-based generators share pair representations, this provides a reusable inference-time control interface. On the complete 82-protein ATLAS test split, MotionPlan reduces trajectory-level C$\alpha$ MAE from 4.60~\AA{} to 3.309~\AA{} on the stride-256, $L=9$ benchmark and improves frame/consecutive-frame Pearson correlation from 0.62/0.53 to 0.67/0.59. On the 100~ns benchmark, it lowers JSD from $0.52$ to $0.44$ and raises state recall from $0.36$ to $0.59$ while retaining comparable molecular quality. These results show that LLM reasoning can translate structural evidence into actionable motion priors for frozen protein dynamics generators.
\end{abstract}

\section{Introduction}
\label{sec:introduction}

\begin{figure*}[t]
\centering
\includegraphics[width=0.90\textwidth]{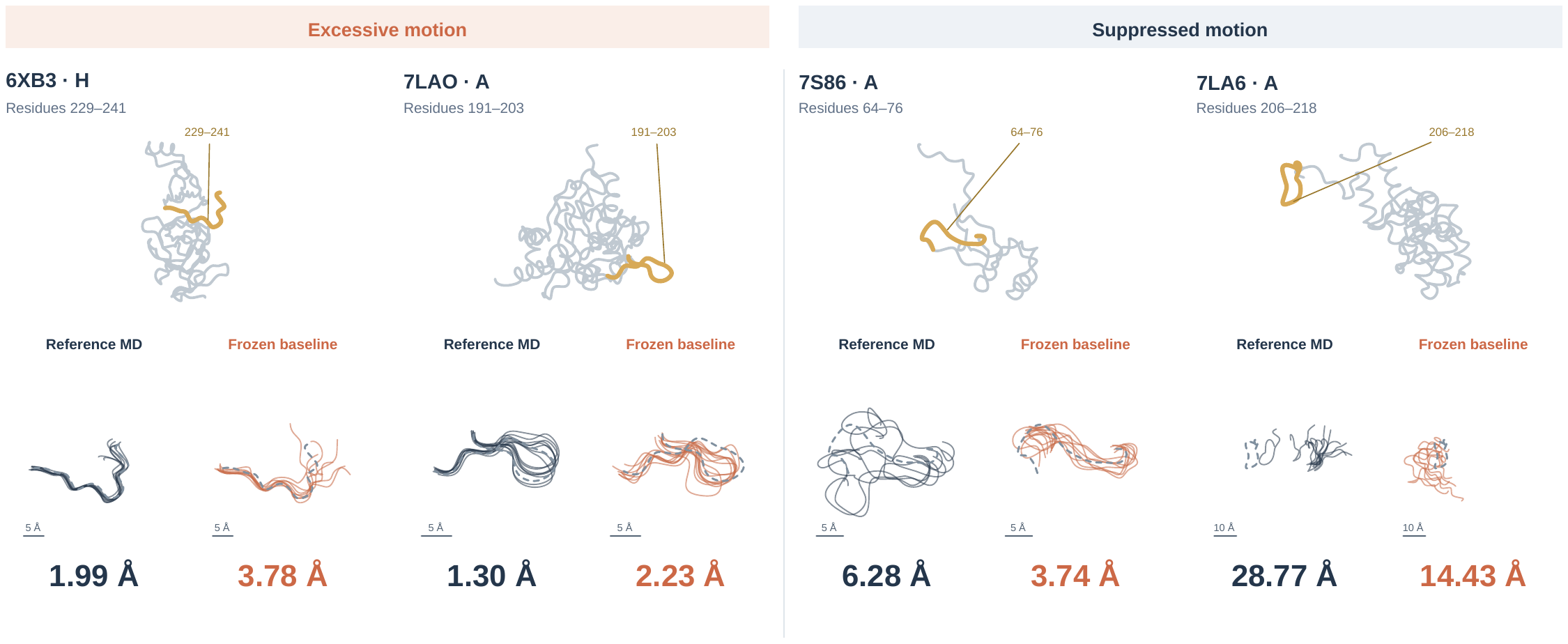}
\caption{Failure cases of a frozen protein dynamics generator. Relative to reference MD, the baseline exaggerates motion in residues 229--241 of 6XB3-H and 191--203 of 7LAO-A, but suppresses motion in residues 64--76 of 7S86-A and 206--218 of 7LA6-A; local RMSDs are annotated in the figure. These cases expose the gap between plausible conformational marginals and the rigid--flexible coordination governing frame-to-frame dynamics.}
\label{fig:hook}
\end{figure*}

Proteins function through conformational ensembles and transitions rather than a single static structure~\citep{henzlerwildman2007dynamic}. Molecular dynamics (MD) resolves these processes atomistically but is costly at long timescales. Diffusion- and flow-based generators amortize this cost by learning conformation distributions~\citep{jing2023eigenfold,zheng2023dig,jing2024alphaflow,wang2024confdiff,lewis2025bioemu,daigavane2025jamun,lu2025eba} or ordered rollouts~\citep{klein2023timewarp,jing2024mdgen,cheng2025alphafolding,costa2025equijump,shen2025confrover,costa2025deepjump,shen2026starmd,liu2026protdyn}. However, their objectives emphasize marginal or short-window statistics rather than the coordination organizing transitions. Rollouts may thus look plausible while coupling flexible loops too tightly to rigid cores or decoupling residues that should move coherently, with errors compounding over time. Figure~\ref{fig:hook} illustrates excessive and suppressed conformational change under a frozen baseline.

Retraining for relational control is unattractive: it requires trajectories annotated with desired rigid cores, independently varying regions, and their boundaries, which MD datasets do not provide, and re-optimizing large generators for each signal is expensive. From a world-model perspective, current generators learn molecular \emph{states} and transitions but expose neither an inference-time \emph{action} for residue coordination nor an \emph{evaluator} that maps structural evidence to that action. Dynamic fidelity is thus inherited from training data and sampling noise.

A shared architectural pattern makes this training-free relational action possible. State-of-the-art systems such as ConfRover and STAR-MD follow a common \emph{Pairformer features + diffusion generation} paradigm. ConfRover couples AlphaFold-derived Pairformer representations with an SE(3)-equivariant diffusion decoder~\citep{shen2025confrover}; STAR-MD retains pretrained pairwise features and explicit pair updates while introducing scalable joint spatiotemporal attention for long-horizon generation~\citep{shen2026starmd}. Across these temporal architectures, pair representations encode residue--residue relations before coordinates are generated. They therefore provide a native intervention surface that expresses coupling and decoupling biases while leaving geometric realization to the coordinate decoder.

\begin{figure*}[t]
\centering
\includegraphics[width=0.85\textwidth]{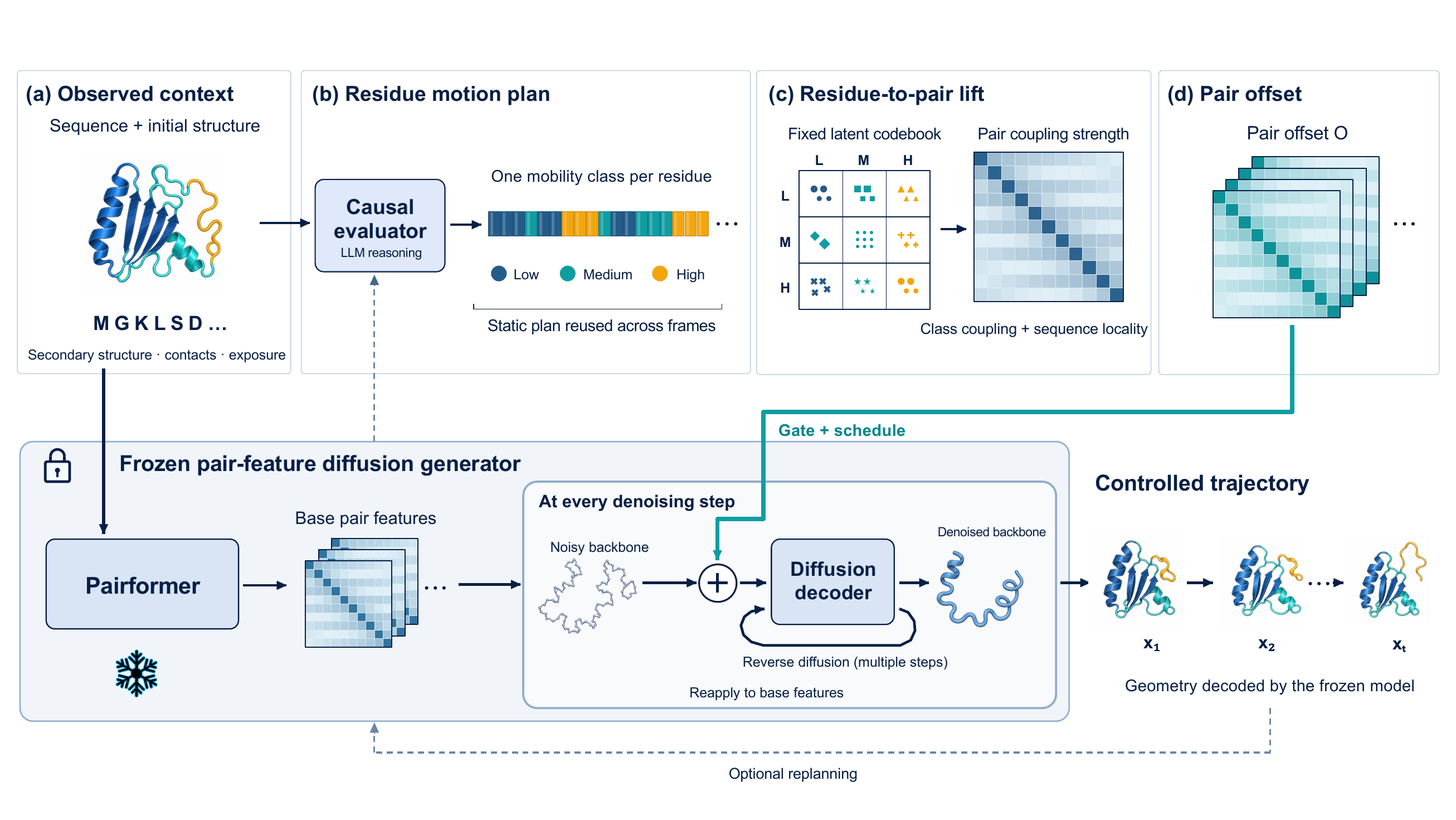}
\caption{MotionPlan overview. (a) A causal evaluator observes sequence, initial structure, secondary structure, contacts, and exposure. (b) The LLM produces a static residue motion plan reused across frames; confidence weighting is omitted for clarity. (c) A fixed latent mapping and sequence locality lift the plan to pairwise coupling strengths. (d) A gate and denoising schedule form pair offset $O$, which is reapplied to the frozen generator's base pair features at every reverse step without accumulation. The frozen Pairformer--diffusion generator decodes the trajectory, with optional replanning when the conditioning context changes.}
\label{fig:method}
\end{figure*}

We introduce \textbf{MotionPlan}, a training-free framework that follows the pipeline summarized in Figure~\ref{fig:method}. A causal LLM evaluator reasons jointly over the observed sequence, initial structure, secondary structure, contacts, exposure, and broader structural context to assign each residue a coordination role and confidence. The resulting static residue plan is lifted into localized pairwise coupling strengths and converted into a gated pair offset. This offset is reapplied to the frozen generator's base pair features at every denoising step, while the pretrained diffusion decoder remains responsible for backbone geometry and the resulting trajectory; optional replanning can refresh the residue plan when the conditioning context changes.

A trajectory-level coupling plan provides a coherent relational prior throughout the rollout. On the stride-256, $L=9$ short-horizon benchmark, MotionPlan reduces C$\alpha$ trajectory-change MAE by 28.1\% relative to ConfRover and improves frame/consecutive-frame Pearson correlation by 8.1\%/11.3\%, respectively. At 100 ns, it reduces JSD by 15.4\% and improves state recall by 63.9\% relative to ConfRover. Relaxed-structure MolProbity remains comparable, differing by only 1.2\% from ConfRover. Together, these results show that controlling inter-residue coordination improves short- and long-horizon dynamics while retaining the baseline level of molecular quality.

Our contributions are threefold:
\begin{itemize}
    \item We identify a relational-control gap in protein dynamics generation: plausible conformational marginals alone leave the inference-time coordination or decoupling of residue pairs unspecified.
    \item We introduce MotionPlan, a training-free LLM-guided mechanism that jointly predicts per-residue coordination roles and five-level ordinal confidence scores, then converts them into a deterministic signed pair-latent offset. The frozen generator retains authority over coordinate-space motion directions and all model parameters remain fixed.
    \item We demonstrate improved short- and long-horizon dynamic fidelity on the complete 82-protein ATLAS test split while maintaining molecular quality at the ConfRover baseline level, establishing pair representations as a reusable control surface for inter-residue coordination.
\end{itemize}

\section{Related Works}
\label{sec:related_work}

\subsection{Protein conformation and trajectory generation}
Protein ensemble generators complement static predictors such as AlphaFold2/3 and ESMFold~\citep{jumper2021alphafold,abramson2024alphafold3,lin2023esmfold}. AlphaFlow adapts structure predictors with flow matching~\citep{jing2024alphaflow}; diffusion and graph methods model equilibrium conformations~\citep{jing2023eigenfold,lu2024str2str,zheng2023dig,wang2024confdiff}; and BioEmu, Energy-based Alignment, and JAMUN improve state coverage or physical weighting~\citep{lewis2025bioemu,lu2025eba,daigavane2025jamun}. De novo backbone generators~\citep{yim2023framediff,lin2023genie,wu2024foldingdiff} likewise produce unordered structures. Trajectory models instead learn temporal transitions through finite-lag propagation~\citep{klein2023timewarp,costa2025equijump,costa2025deepjump}, joint frame modeling~\citep{cheng2025alphafolding,jing2024mdgen}, autoregressive diffusion~\citep{shen2025confrover}, spatiotemporal attention~\citep{shen2026starmd}, or structure tokens~\citep{liu2026protdyn}. MotionPlan is orthogonal: it adds inference-time inter-residue control to a trained generator.

\subsection{Inference-time diffusion guidance}
Diffusion guidance steers a pretrained reverse process with a condition or external objective, building on denoising diffusion models~\citep{ho2020ddpm}. Classifier and classifier-free guidance modify the denoising prediction~\citep{dhariwal2021diffusion,ho2022classifierfree}, while Universal Guidance and diffusion posterior sampling incorporate general objectives or measurement consistency~\citep{bansal2023universal,chung2023dps}. Training-free energy or differentiable-loss guidance further supports plug-and-play control~\citep{yu2023freedom,song2023lossguided}; decoupled annealing and twisted sequential Monte Carlo improve exploration or conditional sampling beyond local gradient corrections~\citep{zhang2025daps,wu2023practical}. In protein generation, Energy-based Alignment uses physical feedback with additional optimization~\citep{lu2025eba}. MotionPlan instead keeps all weights fixed and injects a signed coordination bias directly into the protein pair representation, avoiding class gradients, coordinate-space likelihoods, and training-stage alignment.

\subsection{Protein language models}
Protein language models span large-scale sequence representation~\citep{elnaggar2022prottrans,lin2023esmfold}, structure-aware vocabulary learning~\citep{su2024saprot}, sequence generation, and multimodal sequence--structure--function modeling~\citep{madani2020progen,hayes2025esm3}. Structure-token systems represent conformations discretely~\citep{lu2025slm}; MD-LLM-1 generates tokenized conformational sequences~\citep{murtada2025mdllm}; and ProTDyn extends this paradigm to multi-timescale dynamics~\citep{liu2026protdyn}. MotionPlan uses an LLM differently: as a frozen evaluator that converts residue identity, secondary structure, contacts, and exposure into coordination roles for a continuous-coordinate generator.

\section{Method}
\label{sec:method}

\subsection{Overview and problem formulation}
Let $a=(a_1,\ldots,a_N)$ denote a protein sequence and let $x_t\in\mathbb{R}^{N\times 3}$ denote the C$\alpha$ coordinates of its state at physical time index $t$; the same construction applies when the backbone state contains residue frames or additional atoms. Given an observed state $x_0$, a pretrained dynamics generator samples $X_{1:T}=(x_1,\ldots,x_T)$ from $p_\theta(X_{1:T}\mid a,x_0)$, where all parameters $\theta$ are frozen. MotionPlan exposes an inference-time action over \emph{inter-residue coordination}: it specifies which regions remain tightly coupled, which regions decouple, and which regions mediate their boundary, while the generator's learned geometric prior realizes the corresponding coordinate evolution. Figure~\ref{fig:method} presents this pipeline from observed context and a reusable residue motion plan, through the residue-to-pair lift and gated pair offset, to repeated pair-feature control inside the frozen diffusion generator.

We consider the common pair-feature diffusion interface. At outer trajectory step $t$, the backbone constructs single features $s_t\in\mathbb{R}^{N\times d_s}$, pair features $z_t\in\mathbb{R}^{N\times N\times d_z}$, and an optional temporal state $h_t$. Starting from a noisy structural state $y_t^{(K)}$, its $k$-th reverse-SDE update has the abstract form
\begin{equation}
 y_t^{(k-1)}=\Phi_\theta\!\left(y_t^{(k)},k;s_t,z_t,h_t\right)+\sigma_k\epsilon_k,
 \qquad \epsilon_k\sim\mathcal{N}(0,I),
 \label{eq:base_sampler}
\end{equation}
for $k=K,\ldots,1$. Here $\Phi_\theta$ includes the equivariant coordinate decoder and the base reverse update. The final $y_t^{(0)}$ is decoded as $x_{t+1}$ and incorporated into $h_{t+1}$ when the generator maintains temporal state. MotionPlan supplies a shape-preserving relational offset to $z_t$, while the frozen $\Phi_\theta$ retains coordinate generation and all learned parameters.

\subsection{LLM coordination-role evaluator}
MotionPlan separates \emph{reasoning about coordination} from generating a concrete trajectory. An instruction-following LLM evaluator $E_\psi$ maps the observable sequence and structural context $o_t$ to a residue coordination plan,
\begin{equation}
 M_t=E_\psi(o_t).
 \label{eq:evaluator}
\end{equation}
The observation contains only information available from the conditioning input: amino-acid identity, the initial structure, secondary structure, terminal proximity, contact count and density, graph centrality, burial depth, solvent exposure, and local packing. It never includes a future frame, reference displacement, RMSF, PCA/tICA coordinates, evaluation stride, or any score derived from the target trajectory.

The LLM jointly interprets these cues to assign contextual coordination roles and evidential confidence. Buried, well-connected residues tend to be rigid anchors, exposed loops and termini flexible executors, and their interfaces transitional linkers. Confidence is a five-level ordinal target whose order reflects increasing support. The structured residue-indexed plan is
\begin{equation}
 M_t=\{m_{t,i}\}_{i=1}^{N},\qquad
 m_{t,i}=(c_{t,i},r_{t,i}),
 \label{eq:motion_plan}
\end{equation}
where $c_{t,i}\in\{0,1,2\}$ denotes \textsc{Low}, \textsc{Medium}, or \textsc{High} and $r_{t,i}\in\{1,2,3,4,5\}$ is the emitted ordinal confidence level. Level 1 means weak evidence for the assigned role and level 5 means strong evidence; the output schema therefore contains a discrete five-level score rather than an arbitrary real value. MotionPlan converts this score to an internal amplitude weight only after inference,
\begin{equation}
 \rho_{t,i}=\frac{r_{t,i}-1}{4}\in\left\{0,\frac14,\frac12,\frac34,1\right\}.
 \label{eq:confidence_norm}
\end{equation}
Semantically, \textsc{Low} marks a rigid anchor, \textsc{Medium} a transitional linker, and \textsc{High} a flexible executor. The plan contains exactly one valid role and one integer confidence level for every residue. This joint role--confidence output preserves the ordinal regression semantics at the evaluator interface, while $\rho_{t,i}$ provides a calibrated weight for pair control.

\subsection{From ordinal role confidence to signed pair control}
Pair-feature generators naturally represent residue--residue relations, making them the appropriate control surface for the LLM output. MotionPlan lifts the joint role--confidence plan to an offset $O_t\in\mathbb{R}^{N\times N\times d_z}$ with the same shape as the pair representation. We assign the three coordination roles ordered scalar values
\begin{equation}
 s_0=-1,\qquad s_1=0,\qquad s_2=+1,
 \label{eq:role_scale}
\end{equation}
corresponding to rigid coordination, transition, and flexible decoupling. The order supplies an explicit semantic axis: negative values favor rigid coordination, positive values favor flexible decoupling, and zero is neutral. We use a single fixed unit-norm basis $u\in\mathbb{R}^{d_z}$, selected once with a fixed seed and shared across all proteins and samples. Because every role combination acts along this common axis, the sign and magnitude are determined entirely by the ordered role values and ordinal confidence rather than by a learned or role-specific random codebook.

For a valid-residue mask $b_i\in\{0,1\}$ and sequence-locality factor
\begin{equation}
 \ell_{ij}=\exp\!\left(-\frac{|i-j|}{\max(4,N/4)}\right),
 \label{eq:local_decay}
\end{equation}
define the mean normalized confidence and signed pair coefficient as
\begin{equation}
 \bar\rho_{t,ij}=\frac{\rho_{t,i}+\rho_{t,j}}{2},\qquad
 \alpha_{t,ij}=\frac{s_{c_{t,i}}+s_{c_{t,j}}}{2}\,\bar\rho_{t,ij}.
 \label{eq:pair_coefficient}
\end{equation}
The pair offset is
\begin{equation}
 O_{t,ij}=m\,b_i b_j\,\ell_{ij}\,\alpha_{t,ij}\,u,
 \label{eq:pair_lift}
\end{equation}
where $m$ controls the global intervention strength. For a confident \textsc{Low}--\textsc{Low} pair, $\alpha_{t,ij}=-\bar\rho_{t,ij}$ produces the strongest negative coordination bias; for a confident \textsc{High}--\textsc{High} pair, $\alpha_{t,ij}=+\bar\rho_{t,ij}$ produces the strongest positive decoupling bias. A \textsc{Low}--\textsc{High} pair has $\alpha_{t,ij}=0$, expressing a neutral boundary, while \textsc{Low}--\textsc{Medium} and \textsc{Medium}--\textsc{High} pairs receive half-strength signed biases. \textsc{Medium}--\textsc{Medium} is also neutral. Because both the role average and confidence average are symmetric, $O_{t,ij}=O_{t,ji}$ before any backbone-specific processing. The fixed basis determines the pair-space direction, the role scale determines its sign, and the ordinal confidence controls its magnitude. Sequence locality emphasizes nearby interactions while retaining long-range coordination. The operation costs $O(N^2d_z)$, matching the asymptotic size of the pair tensor already present in the backbone.

\subsection{Repeated pair-latent injection}
At reverse step $k$, MotionPlan presents the controlled pair feature
\begin{equation}
 \widetilde z_t^{(k)}=z_t^{(k)}+\gamma\,q_k\,O_t,
 \label{eq:latent_injection}
\end{equation}
where $\gamma$ is an inference-time gate and $q_k$ controls the injection strength at reverse step $k$. The controlled sampler is therefore
\begin{equation}
 y_t^{(k-1)}=\Phi_\theta\!\left(y_t^{(k)},k;s_t,\widetilde z_t^{(k)},h_t\right)+\sigma_k\epsilon_k.
 \label{eq:controlled_sampler}
\end{equation}
The pair offset is applied at every reverse step. Equation~\ref{eq:latent_injection} re-applies the LLM-inferred signed coupling field to the current base pair state, keeping the confidence-weighted relational bias consistent throughout denoising while avoiding offset accumulation across steps. The injected bias organizes coordination, decoupling, and neutral boundaries; the equivariant decoder produces every coordinate and autonomously realizes conformational change under this controlled pair context.

\begin{table*}[t]
\centering
\scriptsize
\setlength{\tabcolsep}{2.4pt}
\caption{Detailed multi-start evaluation at stride 256 and $L=9$. C$\alpha$ MAEs are in \AA. Correlations are higher-is-better and MAEs are lower-is-better.}
\label{tab:short_s256}
\resizebox{\textwidth}{!}{%
\begin{tabular}{lccccccccccccc}
\toprule
& \multicolumn{3}{c}{C$\alpha$ Pearson $\uparrow$} & \multicolumn{3}{c}{PCA Pearson $\uparrow$} & Diversity & \multicolumn{3}{c}{PCA MAE $\downarrow$} & \multicolumn{3}{c}{C$\alpha$ MAE $\downarrow$} \\
\cmidrule(lr){2-4}\cmidrule(lr){5-7}\cmidrule(lr){8-8}\cmidrule(lr){9-11}\cmidrule(lr){12-14}
Method & Traj. & Frame & $\Delta$ & Traj. & Frame & $\Delta$ & Pair RMSD & Traj. & Frame & $\Delta$ & Traj. & Frame & $\Delta$ \\
\midrule
MDGen & 0.57 & 0.46 & 0.41 & 0.18 & 0.13 & 0.11 & 1.34 & 7.66 & 1.54 & 1.06 & 6.00 & 1.14 & 0.85 \\
MDGen-S256F9 & 0.56 & 0.45 & 0.38 & 0.21 & 0.19 & 0.11 & 1.59 & 6.90 & 1.47 & 1.03 & 5.33 & 1.13 & 0.82 \\
ConfRover & \textbf{0.77} & 0.62 & 0.53 & \textbf{0.75} & 0.50 & 0.44 & 1.78 & 3.91 & \textbf{1.28} & 0.87 & 4.60 & 0.91 & 0.75 \\
\midrule
MotionPlan & 0.75 & \textbf{0.67} & \textbf{0.59} & 0.74 & \textbf{0.61} & \textbf{0.48} & 2.273 & \textbf{3.65} & 1.46 & \textbf{0.81} & \textbf{3.309} & \textbf{0.872} & \textbf{0.675} \\
\bottomrule
\end{tabular}%
}
\end{table*}

\subsection{Training-free portability and implementation scope}
MotionPlan has no optimization objective and updates no generator parameters. Integrating it requires (i) an LLM evaluator that emits the role and five-level ordinal confidence schema in Eq.~\ref{eq:motion_plan}, (ii) access to an $N\times N\times d_z$ pair representation, and (iii) a shape-preserving relational addition before coordinate decoding. The role scale $(s_0,s_1,s_2)=(-1,0,+1)$, confidence normalization, and fixed basis $u$ are deterministic inference rules shared across proteins. For a step-wise Pairformer--diffusion model, the signed offset is applied to the decoder pair latent independently at each outer frame. For a joint spatiotemporal generator, the same residue-to-pair operator can be broadcast or indexed over time before pair-biased temporal blocks. The sign semantics, ordinal confidence weighting, validity masking, and no-accumulation rule remain unchanged. MotionPlan defines its inference-time action entirely through the LLM-inferred role--confidence plan, with the pretrained generator checkpoint kept fixed.

\section{Experiments}
\label{sec:experiments}

\subsection{Experimental setup}
\paragraph{Data and splits.}
The frozen generators retain the training data and checkpoints of their corresponding baseline methods; MotionPlan uses no trajectory training data and updates no model parameters. We evaluate on the public ATLAS benchmark~\citep{vandermeersche2024atlas} using the split and multi-start construction adopted by ConfRover~\citep{shen2025confrover}. The test split contains 82 proteins, each with three 10,000-frame MD replicas saved every 10~ps. For short-horizon evaluation, each sample contains $L=9$ frames. Start indices are drawn from 1,000--7,000 and temporal strides from $\{128,256,512,1024\}$, yielding 2,706 start/stride conditions in the complete protocol. Short- and long-horizon dynamics are evaluated on the complete 82-protein test split. Molecular quality is evaluated separately on the 38 trajectories shared by the forward and interpolation settings, following the experimental setup of ConfRover~\citep{shen2025confrover}; the stride-256 setting is used for controlled comparisons.

\paragraph{Long-horizon protocol.}
Following the public-ATLAS evaluation protocol of STAR-MD~\citep{shen2026starmd}, we evaluate 100~ns trajectories on the 82-chain test split using 80 frames at 1.2~ns per frame. Every method starts from the same equilibrated structure. All trajectories are aligned to the starting frame of the reference MD trajectory by C$\alpha$ superposition before computing coverage and kinetic statistics. Coverage is evaluated only on structurally valid conformations, and tICA is evaluated only on time-lagged pairs for which both frames are valid, using the joint C$\alpha$ and all-atom validity criteria of STAR-MD.

\paragraph{Baselines and implementation.}
Short-horizon baselines include MDGen, MDGen-S256F9, and ConfRover~\citep{jing2024mdgen,shen2025confrover}; the long-horizon comparison includes MD, MDGen, and ConfRover. In every matched controlled run, the checkpoint, conditioning structure, diffusion schedule, sampling horizon, and random seed are fixed. MotionPlan uses the LLM evaluator to infer residue coordination roles and five-level ordinal confidence scores, deterministically converts them into a signed coupling field, and changes only the relational pair representation supplied to the frozen generator. For each seed, MotionPlan and its corresponding baseline use the same initial structure and sampling seed.

\begin{figure*}[t]
\centering
\includegraphics[width=\textwidth]{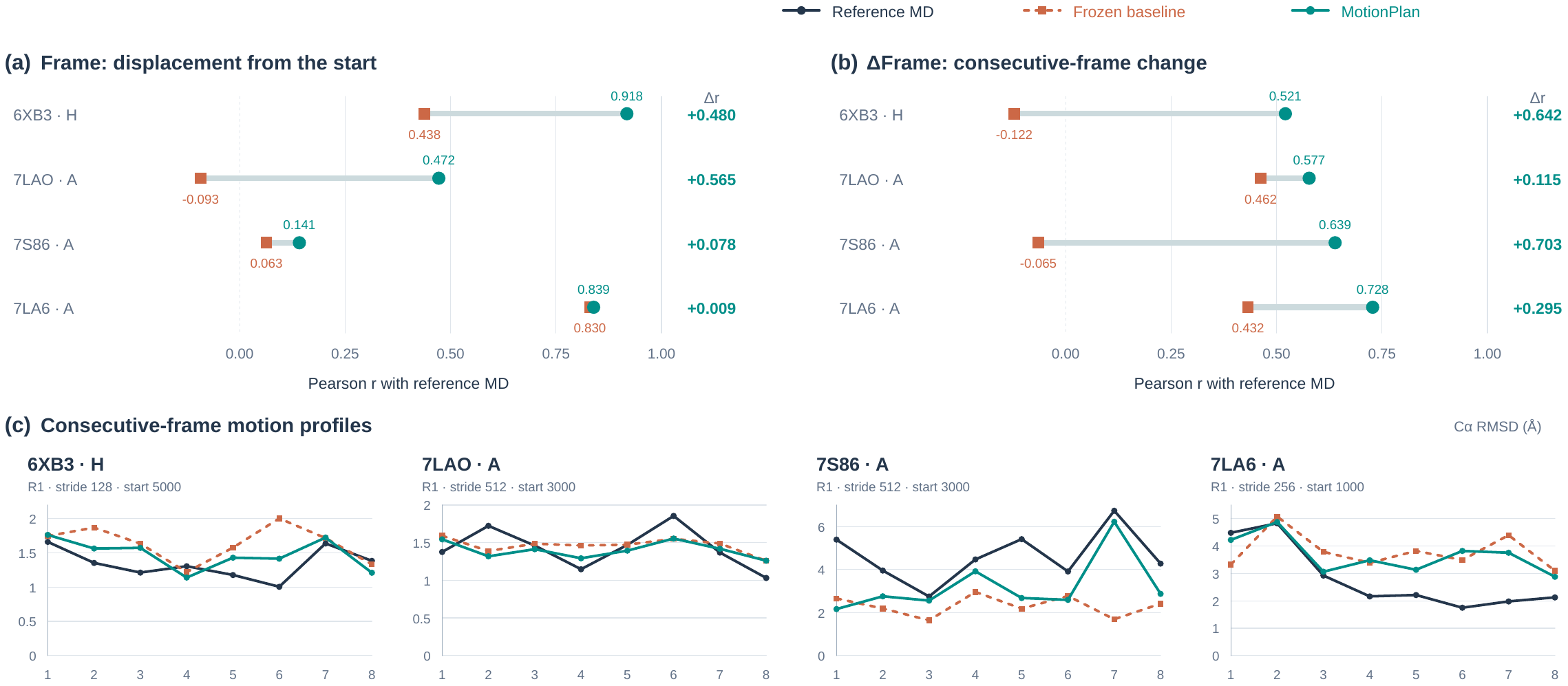}
\caption{Effect of the LLM-inferred signed coupling field on short-horizon fidelity. MotionPlan assigns negative bias to confident rigid--rigid pairs, positive bias to confident flexible--flexible pairs, and neutral bias to rigid--flexible boundaries; five-level ordinal confidence controls the pair-bias magnitude. Panels (a) and (b) compare Pearson correlation for displacement from the initial structure and consecutive-frame change, respectively; the annotated $\Delta r$ values show the gain from the relational intervention. Panel (c) plots consecutive-frame C$\alpha$ RMSD profiles for four representative systems, where MotionPlan better follows the reference peaks and troughs.}
\label{fig:correlation}
\end{figure*}

\begin{table*}[t]
\centering
\scriptsize
\setlength{\tabcolsep}{3.3pt}
\caption{Long-horizon trajectory generation on the public ATLAS 100~ns benchmark with 82 proteins and 80 frames sampled at 1.2~ns intervals, following the evaluation protocol of STAR-MD~\citep{shen2026starmd}. JSD and recall are computed on structurally valid conformations; tICA uses time-lagged pairs with both frames valid. Entries with $\pm$ report the mean and standard deviation across independent random seeds.}
\label{tab:long_horizon}
\resizebox{\textwidth}{!}{%
\begin{tabular}{lcccccc}
\toprule
Method & JSD$\downarrow$ & Recall$\uparrow$ & tICA$\uparrow$ & RMSD $\Delta\downarrow$ & AutoCorr $\Delta\downarrow$ & VAMP-2 $\Delta\downarrow$ \\
\midrule
MD oracle & 0.31 & 0.67 & 0.17 & 0.00 & 0.00 & 0.02 \\
MDGen & $0.56\!\pm\!0.01$ & $0.28\!\pm\!0.01$ & $0.12\!\pm\!0.00$ & $0.38\!\pm\!0.01$ & $0.05\!\pm\!0.00$ & $0.38\!\pm\!0.01$ \\
ConfRover & $0.52\!\pm\!0.01$ & $0.36\!\pm\!0.01$ & $0.15\!\pm\!0.01$ & $0.20\!\pm\!0.00$ & $0.08\!\pm\!0.00$ & $0.47\!\pm\!0.02$ \\
\midrule
MotionPlan & $\mathbf{0.44\!\pm\!0.02}$ & $\mathbf{0.59\!\pm\!0.01}$ & $\mathbf{0.16\!\pm\!0.00}$ & $\mathbf{0.18\!\pm\!0.02}$ & $\mathbf{0.02\!\pm\!0.01}$ & $\mathbf{0.27\!\pm\!0.01}$ \\
\bottomrule
\end{tabular}%
}
\end{table*}

\begin{table*}[t]
\centering
\scriptsize
\setlength{\tabcolsep}{3.8pt}
\caption{Molecular quality of relaxed conformations on the 38 trajectories shared by the forward and interpolation settings, following the experimental setup of ConfRover~\citep{shen2025confrover}. All methods use the same relaxation and geometry-evaluation protocol. Geometric metrics are reported as mean$\pm$standard deviation across the 38 cases; MadraX energy is reported as the mean across these cases.}
\label{tab:molecular_quality}
\resizebox{\textwidth}{!}{%
\begin{tabular}{lccccccc}
\toprule
Method & Rama. outlier\%$\downarrow$ & Rotamer outlier\%$\downarrow$ & Clashscore$\downarrow$ & RMS bonds$\downarrow$ & RMS angles$\downarrow$ & MolProbity$\downarrow$ & MadraX$\downarrow$ \\
\midrule
MD Reference & $\mathbf{0.38\!\pm\!0.49}$ & $\mathbf{1.02\!\pm\!0.89}$ & $\mathbf{0.04\!\pm\!0.16}$ & $\mathbf{0.01\!\pm\!0.00}$ & $\mathbf{1.88\!\pm\!0.05}$ & $\mathbf{0.72\!\pm\!0.18}$ & $-519.3$ \\
MDGen & $0.93\!\pm\!0.86$ & $2.86\!\pm\!1.59$ & $16.14\!\pm\!20.05$ & $0.02\!\pm\!0.02$ & $2.13\!\pm\!0.30$ & $2.24\!\pm\!0.40$ & $-314.7$ \\
ConfRover & $0.58\!\pm\!0.63$ & $1.98\!\pm\!1.48$ & $7.81\!\pm\!6.52$ & $\mathbf{0.01\!\pm\!0.01}$ & $\mathbf{1.88\!\pm\!0.25}$ & $1.72\!\pm\!0.38$ & $\mathbf{-522.2}$ \\
\midrule
MotionPlan & $0.55\!\pm\!0.61$ & $2.06\!\pm\!1.59$ & $8.03\!\pm\!6.34$ & $\mathbf{0.01\!\pm\!0.01}$ & $1.92\!\pm\!0.26$ & $1.70\!\pm\!0.37$ & $-518.7$ \\
\bottomrule
\end{tabular}%
}
\end{table*}

\subsection{Evaluation metrics}
\paragraph{Short-horizon dynamics.}
Following ConfRover, we summarize conformational change at three temporal resolutions: \emph{Trajectory}, the total change accumulated over a rollout; \emph{Frame}, each frame's displacement from the initial structure after alignment; and $\Delta$\emph{Frame}, the displacement between consecutive aligned frames. Pearson correlation is computed across matched start/stride conditions between generated and reference changes in C$\alpha$ coordinate and protein-specific 2D PCA spaces, measuring whether the model preserves their relative variation. MAE averages the corresponding absolute differences and therefore measures calibration in the original displacement scale. Pairwise RMSD is the average aligned RMSD across generated frame pairs and measures whether the rollout retains conformational spread.

\paragraph{Long-horizon dynamics and coverage.}
Following STAR-MD~\citep{shen2026starmd}, Table~\ref{tab:long_horizon} reports Jensen--Shannon distance (JSD) between generated and reference state distributions and reference-state recall in a PCA space fitted to reference MD conformations, with each principal component discretized into 10 evenly spaced bins. Both coverage metrics use only structurally valid generated conformations. The tICA metric measures agreement between per-residue contributions to the leading slow modes of generated and reference trajectories, using only time-lagged pairs whose two frames satisfy the joint C$\alpha$ and all-atom validity criteria. RMSD, autocorrelation, and VAMP-2 deviations are absolute discrepancies from the reference trajectory in structural spread, temporal persistence, and kinetic variational score, respectively~\citep{wu2020variational}.STAR-MD reports CA/AA/CA+AA validity thresholds but not the full C$\alpha$ clash/break implementation, and its evaluation code is unavailable. We therefore omit these metrics to avoid implementation-dependent approximations, while assessing structural validity and quality separately under the ConfRover relaxed-structure protocol in Table~\ref{tab:molecular_quality}.

\paragraph{Molecular quality.}
Following ConfRover's relaxed-structure protocol~\citep{shen2025confrover}, all compared structures are refined with OpenFold~\citep{ahdritz2024openfold} and evaluated under one geometry pipeline. Ramachandran and rotamer outlier rates measure backbone and side-chain torsion violations, clashscore counts severe steric overlaps, and bond/angle RMS values measure covalent-geometry deviations. MolProbity aggregates stereochemical quality~\citep{chen2010molprobity}, while MadraX reports molecular-mechanics energy.

\subsection{Short-horizon Results}
Table~\ref{tab:short_s256} shows that MotionPlan provides the strongest improvement in local temporal organization while preserving trajectory-level coverage. Its signed pair bias gives the frozen generator an explicit relational prior: confident rigid--rigid pairs are encouraged to remain coordinated, flexible--flexible pairs are allowed to decouple, and mixed boundaries are left comparatively neutral. Ordinal confidence further prevents uncertain assignments from dominating the reverse process. The resulting trajectories therefore follow reference frame-to-frame changes more faithfully without collapsing conformational diversity, indicating that the gain comes from better organized motion rather than conservative sampling. Figure~\ref{fig:correlation} shows how this mechanism appears at the system level.

\subsection{Long-horizon Results}
As shown in Table~\ref{tab:long_horizon}, MotionPlan achieves the most balanced long-horizon behavior, recovering broader reference-state support while maintaining the slow collective modes and temporal persistence of the underlying dynamics. Reapplying the signed coupling field throughout denoising prevents the relational prior from being washed out as errors accumulate across a rollout. Rigid regions remain coherent anchors, whereas flexible regions can explore alternative metastable states under the generator's learned coordinate-space dynamics. This separation explains why the method improves both conformational coverage and kinetic fidelity rather than trading one for the other.

\subsection{Molecular Quality}
Table~\ref{tab:molecular_quality} confirms that MotionPlan preserves the relaxed-structure quality of the frozen baseline while improving how conformational changes are organized over time. This behavior follows from applying control in pair-latent space rather than directly perturbing coordinates: the intervention changes which residues move together, while the pretrained equivariant decoder and the shared relaxation procedure continue to enforce local stereochemistry and covalent geometry. The molecular-quality results therefore rule out invalid or highly strained structures as the source of the dynamic gains.

\subsection{Ablation Study}
\begin{table*}[t]
\centering
\small
\setlength{\tabcolsep}{4.0pt}
\caption{Factorized short-horizon ablation for MotionPlan at stride 256 and $L=9$. The three panels separately vary the coupling-plan source, residue-to-pair lift, and injection schedule. All metrics are computed in C$\alpha$ coordinate space; correlations are higher-is-better and MAEs are lower-is-better.}
\label{tab:ablation}
\textbf{(a) Coupling-plan source}\par\vspace{2pt}
\resizebox{\textwidth}{!}{%
\begin{tabular}{lcccc}
\toprule
Variant & Frame $r\uparrow$ & $\Delta$Frame $r\uparrow$ & Traj. MAE$\downarrow$ & $\Delta$Frame MAE$\downarrow$ \\
\midrule
Deterministic proxy evaluator & 0.63 & 0.54 & 4.21 & 0.742 \\
Shuffled role assignment & 0.61 & 0.51 & 4.47 & 0.771 \\
Random role assignment & 0.59 & 0.49 & 4.93 & 0.801 \\
Confidence-free scaling & 0.65 & 0.57 & 3.61 & 0.699 \\
\textbf{MotionPlan} & \textbf{0.67} & \textbf{0.59} & \textbf{3.309} & \textbf{0.675} \\
\bottomrule
\end{tabular}%
}
\vspace{5pt}

\textbf{(b) Residue-to-pair lift}\par\vspace{2pt}
\resizebox{\textwidth}{!}{%
\begin{tabular}{lcccc}
\toprule
Variant & Frame $r\uparrow$ & $\Delta$Frame $r\uparrow$ & Traj. MAE$\downarrow$ & $\Delta$Frame MAE$\downarrow$ \\
\midrule
Unsigned role scale & 0.58 & 0.48 & 5.08 & 0.823 \\
Negated basis ($-u$) & 0.55 & 0.45 & 5.61 & 0.876 \\
No locality decay & 0.65 & 0.56 & 3.72 & 0.706 \\
\textbf{MotionPlan} & \textbf{0.67} & \textbf{0.59} & \textbf{3.309} & \textbf{0.675} \\
\bottomrule
\end{tabular}%
}
\vspace{5pt}

\textbf{(c) Injection schedule}\par\vspace{2pt}
\resizebox{\textwidth}{!}{%
\begin{tabular}{lcccc}
\toprule
Variant & Frame $r\uparrow$ & $\Delta$Frame $r\uparrow$ & Traj. MAE$\downarrow$ & $\Delta$Frame MAE$\downarrow$ \\
\midrule
First-step injection only & 0.62 & 0.53 & 4.57 & 0.754 \\
\textbf{MotionPlan} & \textbf{0.67} & \textbf{0.59} & \textbf{3.309} & \textbf{0.675} \\
\bottomrule
\end{tabular}%
}
\end{table*}

Table~\ref{tab:ablation} isolates each design choice. Panel (a) shows that plan quality matters: shuffling roles while preserving their aggregate distribution already degrades performance, confirming that the gain requires assigning the right coordination intent to the right regions, not just the right role counts; completely random assignment degrades further. Panel (b) shows that the lift must preserve direction and locality: inverting the basis to $-u$ produces the worst result in this panel, as it actively reverses coupling and decoupling throughout the field, while removing the role sign or locality decay each degrades results in smaller but consistent ways. Panel (c) confirms that repeated injection is essential, since a single offset is progressively absorbed by the frozen reverse process.

\section{Conclusion}
\label{sec:conclusion}

Protein dynamics generation requires more than plausible conformations: it must preserve the inter-residue relationships that organize frame-to-frame transitions. Existing generators learn structural statistics but cannot specify residue coordination at inference time, so local errors accumulate and training a separate controller would require paired motion supervision that MD data do not provide. MotionPlan fills this relational-control gap by translating sequence and structural evidence into residue-level intent and injecting it through the pair representations of a frozen generator, while leaving coordinate generation to the pretrained dynamics model. Improvements across short and long horizons show that this prior improves temporal organization rather than merely matching isolated states. Pair representations therefore offer a reusable control interface that can also support future learnable or physics-aware evaluators. We also discuss the current limitations of MotionPlan and promising directions for future work in Appendix~\ref{app:limitations}.

\paragraph{AI use statement.}

Generative AI tools were used to aid and polish writing and to assist with the retrieval and discovery of related work. All AI-assisted content was carefully reviewed, verified, and, where necessary, revised by the authors. In particular, retrieved literature and associated claims were checked against the original sources. We take responsibility for the final content of this work, including text, claims, or artifacts produced with the aid of generative AI.

\paragraph{Reproducibility statement.}

MotionPlan requires no parameter updates; the main text and appendix specify the LLM setup, prompt, pair-latent intervention and evaluation protocol. Code will be released upon acceptance.

\bibliography{iclr2027_conference}
\bibliographystyle{iclr2027_conference}

\def\MotionPlanMainDocument{1}
\ifdefined\MotionPlanMainDocument
  \newcommand{\motionplanmethodref}{Section~\ref{sec:method}}
  \newcommand{\motionplanpairliftref}{Eq.~\ref{eq:pair_lift}}
  \newcommand{\motionplanend}{}
  \clearpage
  \appendix
\else
  \documentclass{article}
  \pdfminorversion=7

  \usepackage{iclr2027_conference,times}
  
  \usepackage{hyperref}
  \usepackage{url}
  \usepackage{booktabs}
  \usepackage{graphicx}
  \usepackage{float}

  \title{MotionPlan: Training-Free LLM-Guided Residue Motion Planning for Protein Dynamics Generation\\Appendix}
  \newcommand{\motionplanmethodref}{the main paper}
  \newcommand{\motionplanpairliftref}{the pair-lift equation in the main paper}
  \newcommand{\motionplanend}{\end{document}}

  \begin{document}
  \maketitle
  \appendix
\fi

\section{LLM Evaluator Selection and Analysis}
\label{app:llm_selection}

\paragraph{Selection and call frequency.}
We use GPT-5.6-sol as the default evaluator because the role assignment requires joint reasoning over sequence identity, secondary structure, contact context, and burial or exposure rather than independent residue classification. The evaluator is executed once for each protein-specific conditioning record, and the resulting ordered residue plan is cached. Each generated trajectory loads this cached plan once and reuses it for all rollout frames and reverse-SDE denoising steps, so the number of LLM calls does not grow with trajectory length, the number of trajectories drawn from that record, or the number of diffusion steps. GPT-5.6-sol provided the strongest short-horizon fidelity in our model-choice study while consistently satisfying the strict output contract.

\begin{table}[H]
\centering
\small
\caption{LLM evaluator choice under the short-horizon protocol at stride 256 and $L=9$. Only the principal C$\alpha$ fidelity metrics are reported. All evaluators use the same prompt, request wrapper, output-validation contract, residue-to-pair lift, and every-step injection schedule.}
\label{tab:llm_choice}
\resizebox{\columnwidth}{!}{%
\begin{tabular}{lccc}
\toprule
Evaluator & Frame $r\uparrow$ & $\Delta$Frame $r\uparrow$ & Traj. MAE$\downarrow$ \\
\midrule
GPT-5.6-sol & \textbf{0.67} & \textbf{0.59} & \textbf{3.309} \\
Opus-4.8 & 0.66 & 0.58 & 3.42 \\
GPT-5.6-luna & 0.65 & 0.56 & 3.64 \\
Seed-2.1-pro & 0.64 & 0.55 & 3.78 \\
DeepSeek-4.0-pro & 0.63 & 0.54 & 4.02 \\
Qwen-7B & 0.60 & 0.50 & 4.72 \\
\bottomrule
\end{tabular}%
}
\end{table}

Table~\ref{tab:llm_choice} shows a monotonic degradation as evaluator capacity decreases. Opus-4.8 is the closest alternative, but GPT-5.6-sol remains best on all three metrics. Smaller evaluators most often confuse flexible loops with neutral boundary residues or assign uniformly high confidence, which weakens the signed pair field. Because the cached plan is reused throughout generation, the selected evaluator adds no per-frame or per-denoising-step LLM latency.

\section{Prompt Template and Output Schema}
\label{app:prompt}

The production prompt is deliberately compact and excludes future coordinates, reference trajectories, force-field energies, and evaluation labels. It presents residues in sequence order and requests one role and one five-level ordinal confidence value per residue. The following pseudocode captures the model-facing contract; implementation-specific transport metadata is omitted.

\begin{minipage}{\linewidth}
\small
\begin{verbatim}
SYSTEM:
  Infer a residue-level coordination plan for a frozen protein-dynamics
  generator. Use only the supplied sequence and structural context.

ROLE SEMANTICS:
  Low    = constrained/rigid residue; supports negative pair bias
  Medium = neutral or rigid-flexible boundary; supports near-zero bias
  High   = flexible/decoupled residue; supports positive pair bias

CONFIDENCE:
  Integer in {1,2,3,4,5}; 1 = weak evidence, 5 = strong evidence.
  Confidence changes bias magnitude, never the meaning of the role.

INPUT:
  sequence: [a_1, ..., a_N]
  residues: for i = 1..N
    {index: i,
     secondary_structure: ss_i,
     contact_context: contacts_i,
     burial_exposure: exposure_i}

OUTPUT (JSON only, sequence order, exactly N entries):
  {"residues": [
    {"index": 1, "role": "Low|Medium|High", "confidence": 1..5},
    ...
  ]}

CONSTRAINTS:
  Do not output coordinates, forces, energies, prose, or extra keys.
  Do not omit, duplicate, or reorder residues.
\end{verbatim}
\end{minipage}

The response is accepted only if it is valid JSON, contains exactly $N$ entries in the input order, uses a role in $\{\textsc{Low},\textsc{Medium},\textsc{High}\}$, and assigns an integer confidence in $\{1,2,3,4,5\}$. Parsing or schema failure is recorded explicitly and the response is excluded rather than repaired with invented residue labels. After validation, confidence is normalized as $\rho_i=(r_i-1)/4$ and the plan is lifted to pair space exactly as defined in \motionplanmethodref.

\section{Limitations and Future Work}
\label{app:limitations}

\paragraph{Limitations.}
The evaluator reasons from sequence and compact structural descriptors rather than an explicit physical force field. It therefore cannot directly test whether a proposed role assignment lowers energy, preserves solvent-mediated interactions, or respects the instantaneous force balance. The three roles and five confidence levels are intentionally coarse, and a plan produced once per protein-specific conditioning record cannot adapt when the generated structure enters a new metastable state. The current lift also projects every scalar coefficient through a fixed pair-space basis $u$; this preserves the frozen-generator interface but cannot learn channel-specific control directions. Finally, the global magnitude $m$, injection strength $\gamma$, and denoising schedule $q_k$ are selected manually and may require recalibration across generators, proteins, or rollout horizons. Dependence on an external LLM also introduces latency and availability concerns; strict schema validation prevents malformed output from entering evaluation, but an unavailable or invalid response leaves that case without an LLM plan. At the time of writing, the training code, model checkpoints, and training data for the current state-of-the-art STAR-MD are not publicly available, which precludes evaluating whether our control mechanism transfers to STAR-MD.

\paragraph{Future work.}
A natural extension is to replace the LLM with a physics-aware learnable evaluator. Such a module could consume local geometry, forces, energy terms, solvent exposure, and temporal state, then predict either the residue plan or the pair control field directly. Joint training with the trajectory generator would turn MotionPlan from a training-free intervention into an end-to-end differentiable coordination controller, supervised by trajectory reconstruction together with energy, geometry, and kinetic-consistency objectives. The pair basis $u$, magnitude $m$, injection strength $\gamma$, and schedule $q_k$ could then be learned rather than fixed. LLM plans may still provide initialization or distillation targets, while the final controller grounds its decisions in physical state and adapts them throughout the rollout.

\section{Implementation and Reproducibility Details}
\label{app:implementation}

\paragraph{Control hyperparameters and denoising schedule.}
The configuration used for the reported ConfRover experiments sets the offset magnitude to $m=0.50$ and the inference-time gate to $\gamma=14$. We use 100 reverse-SDE steps and a constant schedule,
\begin{equation}
q_k=1,\qquad k=1,\ldots,100.
\end{equation}
Thus, the same scaled offset is presented at every reverse step. The implementation forms each controlled pair state from the current unmodified pair latent plus the scaled offset; it does not add the offset to an already controlled state, thereby preventing accumulation across denoising steps. Invalid or padded residues are removed by the outer-product validity mask, and the locality factor is implemented exactly as $\exp[-|i-j|/\max(4,N/4)]$.

\paragraph{Backbone checkpoint and injection site.}
All controlled and uncontrolled ConfRover runs load the same frozen checkpoint, \texttt{confrover\_base\_20m\_v1\_0.pt}. MotionPlan modifies the pair latent only: the shape-preserving offset is added immediately before the reverse-step denoiser consumes that pair state, while the single-residue latent and all model parameters remain unchanged. The same base pair latent is used to reconstruct the controlled input at the next step.

\paragraph{LLM execution and validation.}
The closed-model plan-generation job invokes the model identifier \texttt{gpt-5.6-sol} once per protein-specific conditioning record through the evaluator CLI, with a 900-second per-call timeout and four concurrent workers. The wrapper does not set temperature, top-$p$, or an output-token cap, so these values are inherited from the serving backend rather than claimed as paper-controlled hyperparameters. It extracts the final JSON array, normalizes role strings, validates the role and confidence fields, and accepts a plan only when it contains exactly one entry for every residue in strictly increasing index order.

\motionplanend

\end{document}